\documentclass[twocolumn,showpacs,pra,aps,superscriptaddress]{revtex4}
\usepackage{array}
\usepackage{amsmath}
\usepackage{graphicx}

\begin{document}

\title{Experimentally realizable 
  characterizations of continuous variable Gaussian states}

\author{M. S. Kim}

\author{Jinhyoung Lee}

\affiliation{School of Mathematics and Physics, Queen's University,
  Belfast BT7 1NN, United Kingdom}

\author{W. J. Munro}

\affiliation{Hewlett-Packard Laboratories, Filton Road, Stoke Giord,
  Bristol BS34 8QZ, UK}

\date{\today}

\begin{abstract}
  Measures of entanglement, fidelity and purity are basic yardsticks in
  quantum information processing.  We propose how to implement these
  measures using linear devices and homodyne detectors for continuous
  variable Gaussian states.  In particular, the test of entanglement
  becomes simple with some prior knowledge which 
  is relevant to current experiments. 
\end{abstract}

\maketitle

In the current development of quantum information processing, we have
witnessed the importance of fidelity, entanglement and mixedness.  In
this paper, we propose feasible experimental schemes to measure these
critical quantities for Gaussian continuous-variable systems.  The
proposed schemes require standard laboratory devices such as beam
splitters and phase shifters and highly efficient homodyne detectors.

Entanglement has been mainly confined to theoretical discussions.  Only
very recently, Horodecki and Ekert \cite{Ekert} investigated how to
measure entanglement.  Experimental studies on quantum correlation
between two or more particles have been concentrated on tests of the
Einstein-Podolsky-Rosen (EPR) paradox and Bell's inequality because an
experimental measure of entanglement was not so clear. The present paper 
proposes an experimental scheme to measure a degree of entanglement with
some prior knowledge of the given state.  There are some indirect ways
to test entanglement, for example, by proving the fidelity
higher than the classical limit in quantum teleportation \cite{Braunstein}.  However, this
provides only a sufficient condition for entanglement.  On top of that,
how to measure the fidelity has not been
thoroughly investigated.  Here, we also propose an experiment
to measure how close two states are.

Experiments on continuous-variable quantum information processing have
been concentrated on Gaussian states \cite{Braunstein,Walmsley,Leuchs}.
This is because, due to extremely low efficiency and high dispersion in
high-order nonlinear interaction, experiments have been based on linear
transformation of fields initially in thermal equilibrium.  In thermal
equilibrium, a field is in a Gaussian thermal state \cite{Barnett97} and
the linear transformation keeps the Gaussian nature \cite{Agarwal}.  The
linear transformation of an input field to an output fields is due to a
Hamiltonian composed of linear and/or quadratic bosonic operators $\hat
a$ and $\hat a^\dag$.  Any linear transformation for two-mode fields can
thus be represented by a product of
single-mode squeezing, rotation and displacement operations and two-mode
squeezing and beam-splitting operations \cite{Agarwal}.

To measure how close a quantum state 
$\hat\rho_1$ is to a reference pure state $|\psi\rangle_2$, the
fidelity is defined as ${\cal F}={}_2\langle|\hat\rho_1|\psi\rangle_2$.  
The fidelity is important, for example, to find how
successfully a state is reproduced after a set of local quantum operations 
and classical communications such as a
teleportation process.  However this theoretical concept has not been
thoroughly investigated by experiment.  We propose a feasible
experimental scheme to realize the measurement
of the fidelity.

It is convenient to work with the Weyl characteristic function defined
as $C_i({\bf x})=\mbox{Tr}\hat{\rho}_i\hat{D}_i({\bf x})$.  Here,
the displacement operator is defined as $\hat{D}_i({\bf x})=
\exp(i\hat{{\bf x}}_i{\bf x}^T)$, where the operator
vector $\hat{{\bf x}}_i=(\hat{q}_i, \hat{p}_i)$ with quadrature
operators $\hat{q}_i=(\hat{a}_i+\hat{a}^\dag_i)$ and
$\hat{p}_i=i(\hat{a}^\dag_i-\hat{a}_i)$ and the coordinate 
vector ${\bf x}=(p,-q)$.  The fidelity $\cal F$ is equivalent to the
overlap between the characteristic functions:
\begin{equation}
\label{fidelity-char}
{\cal F}=\frac{1}{\pi}\int d{\bf x} C_1({\bf x}) C_2(-{\bf x}).
\end{equation}
Throughout the paper, matrices are represented in bold face
and operators with hats.  In order to measure the fidelity
of two fields, they are mixed at a beam splitter whose
action is described by
$\hat{B}_{12}=\exp[\theta(\hat{a}_1^\dag\hat{a}_2-\hat{a}_1\hat{a}^\dag_2)]$,
where the reflectivity $r$ and transitivity $t$ of the beam splitter are
determined by $\theta$: $r=\sin\theta$ and $t=\cos\theta$.  The
characteristic function for an output field is
\begin{equation}
\label{char-output}
C_{out}({\bf x})=\mbox{Tr}_2\hat{D}_2({\bf
  x})\mbox{Tr}_1\hat{B}_{12}\hat{\rho}_1\hat{\rho}_2=C_1(r{\bf
  x})C_2(t{\bf x}). 
\end{equation}
By Fourier transforming the characteristic function, the Wigner
function, $W(\alpha)$, is obtained \cite{Barnett97}.  Thus we can easily
see that the Wigner function of the output field at the origin ($\alpha=0$) of 
phase space is directly related to the fidelity:
\begin{equation}
\label{fidelity-wigner}
{\cal F}=\frac{1}{2\pi}\int d{\bf x} C_{out}({\bf x})=2 \pi W(0)
\end{equation}
when $r=-t=1/\sqrt{2}$.  We have found that, after mixing the two fields
at a 50:50 beam splitter, we measure the Wigner function at the origin
of the phase space for one of the output fields, to find how close the
two input fields are.  The Wigner function of a given field can be
measured using optical tomography \cite{Vogel}, which requires some
numerical processes on experimental data. However, if both the input
fields are Gaussian, as the output field is also Gaussian, $W(0)$ is
easily measured using a highly efficient homodyne detector.

The characteristic function of any single-mode Gaussian field is written
as
\begin{equation}
\label{char-single}
C({\bf x})=\exp(-\frac{1}{2}{\bf x}{\bf V}_s{\bf x}^T + i {\bf d}{\bf x}^T)
\end{equation}
where ${\bf V}_s$ is the $2\times 2$ variance matrix defined as
$(V_s)_{ij}=\frac{1}{2}(\langle \{\hat{x}_i, \hat{x}_j\}\rangle)
-\langle \hat{x}_i\rangle\langle\hat{x}_j\rangle$ and ${\bf d}$ is the
displacement vector $d_i=\langle \hat{x}_i\rangle$.  The
quadrature variables are measured by a balanced homodyne detector
\cite{Yuen}, which is a well-known device to detect phase-dependent
properties of an optical field.  The operational representation of the
balanced homodyne detector is $
\hat{O}_{HD}=\hat{q}\cos\chi+i\hat{p}\sin\chi $, where $\chi$ depends on
the local-oscillator phase.  The value $W(0)$ of the Wigner function is
rotationally invariant so the numerical process  becomes 
simpler as we rotate the output field to diagonalize the
variance matrix ${\bf V}_s$.  This can be done by placing an optical
phase shifter before the homodyne detector.  When the variance
matrix is diagonalized the uncertainty $\Delta q\Delta p$ is minimized
and the fidelity (\ref{fidelity-wigner}) is
\begin{equation}
\label{fidelity}
{\cal F}= \frac{1}{\Delta q\Delta
  p}\exp\left[-\frac{1}{2} \left(\frac{\langle \hat q\rangle^2}{\Delta q^2} +  
\frac{\langle \hat p\rangle^2}{\Delta p^2}\right)\right],
\end{equation}
where $\langle\hat p\rangle$, $\langle\hat q\rangle$,
$\langle\hat{p}^2\rangle$ and $\langle\hat{q}^2\rangle$ are measured by a homodyne detector.
We have shown how the fidelity for two Gaussian fields is measured using a homodyne
detector and a beam splitter.  In relation to the two input fields, $\langle
\hat{p}\rangle$ and $\langle\hat q\rangle$ are  proportional to 
the displacement difference between them and $\Delta q
\Delta p$ is the uncertainty over their averaged variance
matrix. When the displacement difference for the two fields is zero, the
fidelity is determined by their average uncertainty: ${\cal F}=1/\Delta p\Delta q$.

We have proposed an experimental scheme to measure the fidelity of two
single-mode fields.  We are now interested in how to measure the 
a degree of entanglement for a two-mode Gaussian field. In
the discussion, we will be able to show how the entanglement measure is
related to the mixedness of the system.  The entanglement of a two-mode
state does not change by displacement operation \cite{Kim02}
so we write the general form of the characteristic function, $C({\bf
  x})$, for a two-mode Gaussian state without the linear displacement
term in Eq.~(\ref{char-single}): $C({\bf x})=\exp(-{\bf x}{\bf V}{\bf
  x}^T/2)$, where ${\bf x}$ is now $(q_1,p_1,q_2,p_2)$ and ${\bf V}$ is
a $4\times 4$ variance matrix for modes 1 and 2.  The variance matrix
can in fact be written using $2\time 2$ block matrices ${\bf L}_1$ and
${\bf L}_2$ for local quadrature variables and ${\bf C}$ and its
transpose ${\bf C^T}$ representing inter-mode correlation,
\begin{equation}
\label{block} {\bf V}=
\begin{pmatrix}
  {\bf L}_1 & {\bf C} \\
  {\bf C^T}&{\bf L}_2 
\end{pmatrix}.
\end{equation}

What are the possible types of entangled continuous-variable states one
can produce?  We start with two independent fields of modes 1 and 2,
which are in thermal equilibrium at temperatures $T_1$ and $T_2$,
respectively.  The variance matrix has only local elements, ${\bf
  L}_1=\tilde{n}_1\openone$ and ${\bf L}_2=\tilde{n}_2\openone$, where
$\openone$ is the $2\times 2$ unit matrix and
$\tilde{n}_i=1+2/[\exp(\hbar\omega/k_BT_i)-1]$ with the Boltzmann
constant $k_B$.  The basic components of a linear transformation are
single-mode squeezing, ${\bf S}_i(s)$, and rotation, ${\bf R}_i(\phi)$,
and two-mode squeezing, ${\bf S}_{12}(s)$,
and beam-splitting, ${\bf B}_{12}$, whose actions are described by their
transformations of the variance matrix:
\begin{eqnarray}
{\bf S}_i (s)=
\begin{pmatrix}
  \mbox{e}^{-s} & 0 \\
  0 & \mbox{e}^s
\end{pmatrix},~~
{\bf R}_i (\phi)=
\begin{pmatrix}
  \cos\phi & \sin\phi \\
  -\sin\phi & \cos\phi
\end{pmatrix},
\nonumber 
\\
{\bf S}_{12}(s) =
\begin{pmatrix}
  \cosh s \openone & \sinh s {\boldsymbol \sigma}_z \\
  \sinh s {\boldsymbol \sigma}_z & \cosh s\openone 
\end{pmatrix},~
{\bf B}_{12}=
\begin{pmatrix}
  t\openone & -r\openone \\
  r\openone &  t\openone 
\end{pmatrix}
\label{transformation}
\end{eqnarray}
where ${\boldsymbol \sigma}_z$ is a Pauli spin matrix.  Any combination,  $\bf U$, 
of these matrices, transforms the variance matrix $\bf V$
into ${\bf U}^T{\bf V}{\bf U}$.  Here,
displacement has not been considered because it does not affect
entanglement.

The degree, $s$, of squeezing normally determines the optimum
entanglement for a given setup \cite{Kim02}.  As squeezing is due to
$\chi^{(2)}$-nonlinear interaction, the efficiency is relatively low and
a combination of squeezers is not practical \cite{Loudon}.  Entangled
states have been produced using two-mode squeezing ${\bf S}_{12}$
\cite{Braunstein,Walmsley} or the 
combination of ${\bf B}_{12}{\bf S}_1(r){\bf S}_2(-r)$ \cite{Leuchs}.
For these linear operations, the variance matrix is represented by
diagonal ${\bf L}_1$, ${\bf L}_2$ and ${\bf C}$.  In particular, if the
two modes are initially in thermal equilibrium at the same temperature
before a transformation, the local matrices ${\bf L}_1$ and ${\bf L}_2$
are identical, for which case we show that the degree of entanglement
becomes simple and can be measured using joint homodyne detectors.  If
such a Gaussian field is decohered in two-mode thermal bath of which the two
modes are in the same temperature, the decohered state is still
represented by the same form of variance matrix \cite{LeeKim}.

The inseparability of a continuous variable Gaussian state is determined
by Peres-Horodecki condition \cite{Simon}. A density operator
$\hat{\rho}$ is entangled if and only if the partially transposed
density operator $\hat{\rho}^{T_2}$ has any negative eigenvalues.  In
order to quantify entanglement, we define the degree of entanglement,
${\cal E}(\hat{\rho})$, as the absolute sum of the negative eigenvalues
of the partially transposed density operator: ${\cal E}(\hat{\rho}) =
\mbox{Tr} |\hat{\rho}^{T_2}| - 1$.  This is an entanglement monotone
\cite{Lee00}.  In the following, we show how ${\cal E}(\hat{\rho})$ is
related to homodyne measurements.

{\em Lemma 1.} -- If the block matrices ${\bf L}_1={\bf L}_2$ and ${\bf
  C}$ are diagonal, {\em i.e.}, the variance matrix of a Gaussian
continuous-variable state has the following form:
\begin{equation}
{\bf V}_0=
\begin{pmatrix}
  n_1 & 0 & c_1 & 0 \\
  0 & n_2 & 0 & c_2 \\
  c_1 & 0 & n_1 & 0 \\
  0 & c_2 & 0 & n_2
\end{pmatrix}
\label{standard-0}
\end{equation}
where $n_1$ or $n_2$ may be smaller than the vacuum limit 1. The degree
of entanglement is given by
\begin{equation}
\label{eq:en}
{\cal E}(\hat{\rho}) = \max \{0,(\delta_1\delta_2)^{-1} -1\} 
\end{equation}
where $\delta_i=n_i-|c_i|$ for $i=1,2$.

We briefly sketch the proof. The main task is to calculate the trace of
$|\hat{\rho}^{T_2}|$ which is the positive operator satisfying
$|\hat{\rho}^{T_2}|^2 = (\hat{\rho}^{T_2})^2$. In order to find the
bound operator $|\hat{\rho}^{T_2}|$ and its trace, all operators are
represented by their characteristic functions respectively, based on the
one-to-one correspondence principle between a bound operator and its
characteristic function \cite{Barnett97}. Then the operator equation,
$|\hat{\rho}^{T_2}|^2=(\hat{\rho}^{T_2})^2$, is converted into an
equation for their corresponding characteristic functions. The equality
of the two Gaussian characteristic functions implies that their variance
matrices and normalization values are the same. Let ${\bf V}_p$ and
$N_p$ be the variance matrix and the normalization value for
$|\hat{\rho}^{T_2}|$.  The unit trace of $\hat{\rho}^{T_2}$ leads its
normalization value to be unity:
$\tilde{N}\equiv\mbox{Tr}\hat{\rho}^{T_2}=1$. Based on the observation
that the transposition is momentum reversal, the variance matrix
$\tilde{\bf V}_0$ of $\hat{\rho}^{T_2}$ is obtained from ${\bf V}_0$
with $\hat{p}_2 \rightarrow -\hat{p}_2$. One thus has the two equations
for ${\bf V}_p$ and $N_p$ as
\begin{eqnarray}
\label{eq:pc}
 & & {\bf V}_p - {\boldsymbol \Omega} {\bf V}_p^{-1} {\boldsymbol \Omega}^T
  = \tilde{\bf V}_0 - {\boldsymbol \Omega} \tilde{\bf V}_0^{-1} {\boldsymbol
  \Omega}^T \\
\label{eq:np}
 & & N_p = \sqrt{\frac{\mbox{det} {\bf V}_p}{\mbox{det} \tilde{\bf V_0}}}.
\end{eqnarray}
In addition, because $|\hat{\rho}^{T_2}|$ is positive, its variance
matrix ${\bf V}_p$ satisfies the uncertainty principle \cite{Simon},
\begin{eqnarray}
\label{eq:up}
  {\bf V}_p + i {\boldsymbol \Omega} \ge 0.
\end{eqnarray}
where ${\boldsymbol \Omega}=i {\boldsymbol \sigma}_y\otimes\openone$
with the Pauli spin matrix ${\boldsymbol \sigma_y}$.  From the solution
${\bf V}_p$ satisfying both Eqs.~(\ref{eq:pc}) and (\ref{eq:up}), $N_p$
in Eq.~(\ref{eq:np}) is obtained. Finally, the degree of entanglement
${\cal E} = N_p - 1$ as $\mbox{Tr}|\hat{\rho}^{T_2}|=N_p$ and
${\cal E}$ is given in particular by Eq.~(\ref{eq:en}) for ${\bf V}_0$.
If and only if the partial transposed density operator is positive
satisfying the uncertainty principle (\ref{eq:up}), ${\bf V}_p =
\tilde{\bf V}_0$ and the state is separable with ${\cal E}=0$. 

For the two-mode vacuum, $n_1=n_2=1$ and $c_1=c_2=0$ so that
$(\delta_i)_0=1$ where $({\boldsymbol\cdot})_0$ denotes the value for the vacuum.
Recalling the definition of the variance matrix,
$\delta_1=\frac{1}{2}(\langle q_1^2\rangle + \langle q_2^2\rangle -
2|\langle q_1q_2\rangle|)$ and $\delta_2=\frac{1}{2}(\langle
p_1^2\rangle + \langle p_1^2\rangle - 2|\langle p_1p_2\rangle|)$, which
can be measured by joint homodyne detectors for modes 1 and 2.
According to Eq.~(\ref{eq:en}) the state is separable with ${\cal E}=0$
when $\delta_1 \delta_2 \ge (\delta_1 \delta_2)_0$.  Otherwise, the
degree of entanglement is ${\cal E}(\hat{\rho}) =
(\delta_1\delta_2)_0/(\delta_1\delta_2)-1$.  We have found that the
entanglement of a Gaussian field in the form (\ref{standard-0}) can be
tested by comparing the joint quadrature variance of the given field
with that of the vacuum.

Let us consider the purity of a state. A degree of purity can be defined
as $P=\mbox{Tr} \hat{\rho}^2=(1/2\pi)\int d{\bf x} C({\bf x}) C(-{\bf
  x})$, where $C({\bf x})$ is the characteristic function of the given
state.  When $P=1$ the state is pure.  The purity of a Gaussian state
with its variance matrix ${\bf V}_0$ is easily calculated using the
determinant of ${\bf V}_0$: $P= 1 / \sqrt{\det {\bf V}_0}$.  We have
already seen that the matrix elements of $V_0$ can be measured by
homodyne detectors so the purity of the Gaussian state can be measured.
To find the relation between the purity and entanglement, we define the
mixedness $M$ of a system: $M=1-P$, which is zero when the system is
pure and becomes unity when it is totally mixed.  According to Lemma 1,
a Gaussian state of the variance matrix ${\bf V}_0$ is separable when
$\delta_1\delta_2-1\geq 0$.  Multiplying a positive quantity
$(n_1+|c_1|)(n_2+|c_2|)-1$ to the both sides of this inequality, we find
the necessary and sufficient condition for separability: $M_{12}-M_1-M_2\geq
2|c_1c_2|$, where $M_1$ and $M_2$ are the mixedness of the field in mode
1 and 2 and $M_{12}$ is that of the whole field. For a pure
two-mode state with $M_{12}=0$, the state is separable iff $M_1=0$ and
$M_2=0$ with $|c_1c_2|=0$.

\begin{figure}[t]
  \centering \includegraphics[width=0.4\textwidth,width=0.3\textwidth]{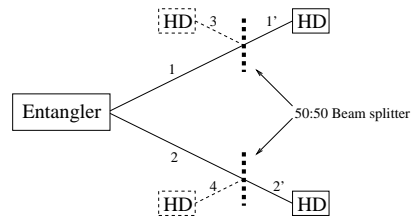}
  \caption{\small Configuration of the setup to test
    entanglement. The dotted devices measure off-diagonal terms of the
    local variance matrices ${\bf L_1}$ and ${\bf L_2}$. The boxed
    devices denoted by HD are homodyne detectors. The numbers refer to
    the modes.}
  \label{fig:beam-splitter}
\end{figure}

So far we have assumed that the block matrices ${\bf L}_1$ and ${\bf
  L}_2$ and ${\bf C}$ are diagonal because this case is relevant to the
current experimental conditions and the degree of entanglement becomes
extremely simple.  However, how do we make sure that there is vanishing
off-diagonal matrix elements?  Measuring the off-diagonal terms of local
matrices is troublesome as it involves the joint measurement of two
quadrature variables for a single mode.  To measure the off-diagonal
elements of ${\bf L}_1$ \cite{heterodyne}, we put a 50:50 beam splitter
which splits the field in mode 1 as schematically shown in
Fig.~\ref{fig:beam-splitter}. Using the beam splitter transformation in
(\ref{transformation}) with $r=t=1/\sqrt{2}$, we find that the field for
three modes $1^\prime, 2$ and 3 is still Gaussian and its variance
matrix is written as
\begin{equation}
\label{block-bs} \frac{1}{2}
\begin{pmatrix}
  {\bf L_1}+\openone &\sqrt{2}{\bf C} &  -{\bf L_1}+\openone \\
  \sqrt{2}{\bf C^T}&{\bf L_2} & -\sqrt{2}{\bf C^T} \\
  -{\bf L_1}+\openone & -\sqrt{2}{\bf C} &  {\bf L_1}+\openone
\end{pmatrix},
\end{equation}
where the unit matrix $\openone$ is due to the vacuum injected into the
unused port of the beam splitter. Now the off-diagonal elements of ${\bf
  L_1}$ can be measured by inter-mode correlation between modes
$1^\prime$ and 3: The mean value of the joint measurement $\langle
q_{1^\prime}p_3\rangle=-V_{12}/2$. Similarly other off-diagonal terms of
the local variance matrices can be obtained.

In fact, we have shown how to find all the matrix elements of the
variance matrix for a Gaussian field so that it is possible to test
entanglement not only for the fields in the form (\ref{standard-0}) but
also for any Gaussian field if the detection efficiency is unity. In
this case we need to generalize the expression (\ref{eq:en}) for the degree of
entanglement which is rather straightforward using the same argument to derive
(\ref{eq:en}). An equivalent but alternative
approach was suggested by Vidal and Werner \cite{Lee00}.

A homodyne detector is composed of two photodetectors. Inefficient
photodetectors may miss photons to detect and reduces the
quantum correlation between two modes.  The detection efficiency thus
determines the feasibility of the proposed schemes, in particular, to
test entanglement of the field. If the efficiencies of the
photodetectors are same, homodyne measurement by imperfect detectors is
equivalent to homodyne measurement by perfect detectors following a beam
splitter, one input port of which is fed by the field to be measured and
the other by the vacuum \cite{Leonhardt94}.  Here, we assume non-unit
detection efficiency due only to missing photons to detect. The
efficiency $\eta$ of the homodyne detector determines the transmission
coefficient $\sqrt{\eta}$ of the beam splitter.  In fact the fictitious
beam splitter affects the testing field as though it is decohered in the
vacuum reservoir. The detection efficiency, assumed the same for the
both homodyne detectors, effectively changes the variance matrix from
${\bf V}$ to ${\bf V}^\prime=\eta {\bf V} + (1-\eta) \openone$.  This is
what is measured by imperfect homodyne detectors.  If ${\bf V}$ is in
the form ${\bf V}_0$, the variance matrix ${\bf V}_0^\prime$ takes the
same form as ${\bf V}_0$ but with modified matrix elements
$n^\prime_i=\eta n_i + (1-\eta)$ and $|c^\prime_i|=\eta |c_i|$ for each
$i$.

\begin{figure}[t]
  \centering \includegraphics[width=0.23\textwidth]{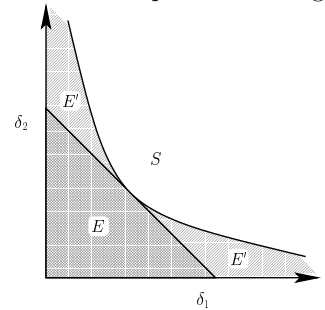}
  \caption{\small Gaussian states with the symmetric variance matrix ${\bf V}_0$ in
    Eq.~(\ref{standard-0}).  Separable states are denoted by $S$ and
    entangled states by $E$ and $E^\prime$ with the boundary of
    $\delta_1 \delta_2 = 1$. Further the two regions $E$ and $E'$ are
    separated by the line $\delta_1+\delta_2=2$. The entanglement
    imposed on a state of the region $E'$ may fail the entanglement test
    due to inefficient detection.  However, the entangled state of the
    region $E$ is always found to be entangled regardless of the
    detection efficiency.}
  \label{fig:ead}
\end{figure}

Consider the effect of the detection efficiency on the inseparability of
the testing fields.  Substituting $n_i^\prime$ and $c_i^\prime$ into the
separability condition, $\delta_1\delta_2\geq 1$, for inefficient
detection, we find a state to be entangled when
\begin{eqnarray}
  \label{eq:deis}
  (n^\prime-|c_1^\prime|)(n_2^\prime-|c_2^\prime|)
  =\eta^2(\delta_1\delta_2-\delta_1-\delta_2+1)
  \nonumber \\
  +\eta(\delta_1+\delta_2-2)+1<1.
\end{eqnarray}
Rearranging this equation, we can easily find that when the original
testing field is characterized by $\delta_1+\delta_2<2$, it is always
found to be entangled regardless of the detection efficiency unless the
efficiency is zero.  Fig.~{\ref{fig:ead}} presents the sets of Gaussian
states on the space of $\delta_1$ and $\delta_2$ where separable states
with $\delta_1 \delta_2 \ge 1$ are denoted by $S$ and entangled states
with $\delta_1 \delta_2 < 1$ by $E+E'$.  All entangled states in the
region $E$ with the condition, $\delta_1 + \delta_2 < 2$, will violate
the inequality, $\delta_1\delta_2\geq 1$, unless the detection
efficiency is zero while some entangled states in $E'$ fail the test of
entanglement.

Reid and Drummond derived the inequality for the quantum correlation
between two mode fields along the line with the EPR argument \cite{Reid,Leuchs}. 
They introduced the uncertainty $V_1$
($V_2$) between the observable $q_1$ ($p_1$) in one mode and
$q_1^\prime$ ($p_1^\prime$) inferred from the observation of the other
mode. Quantum correlation may lead the product of the uncertainties to
be less than the vacuum limit, resulting in the inequality of $V_1 V_2 <
1$. In our notation this inequality can be written as
\begin{eqnarray}
  \label{eq:rdi}
 \delta_1\delta_2 < \frac{n_1 n_2}{(n_1+|c_1|)(n_2+|c_2|)}.
\end{eqnarray}
Note that the right hand side of the inequality is always less than
unity. Thus, the inequality (\ref{eq:rdi}) is sufficient to satisfy our
inseparable condition (\ref{eq:en}) \cite{Ralph}. However, the converse
statement does not hold in general.

We have proposed experimental schemes to measure the fidelity, the
purity and the degree of entanglement of a given system with some prior
knowledge.  The scheme consists of beam splitters and balanced homodyne
detectors which are well-established experimental tools to study quantum
optics.  When the detection efficiency is unity, the scheme can be
extended to a general two-mode Gaussian state to test its entanglement.

We thank Prof. G. J. Milburn and Dr. M. Plenio for discussions and the
UK EPSRC.


\begin{references}
\bibitem{heterodyne} This can also be measured using a heterodyne
  detector.
\end{references}
\end{document}